%% file: 00_main.tex
\documentclass[
twocolumn,
]{ceurart}

\usepackage{listings}
\usepackage{amssymb}
\usepackage{natbib}
\usepackage{geometry}
\usepackage{subfiles}
\usepackage{graphicx}
\usepackage{fontawesome5}
\graphicspath{ {figures/} }

\begin{document}


\copyrightyear{2023}
\copyrightclause{Copyright for this paper by its authors. Use permitted under Creative Commons License Attribution 4.0 International (CC BY 4.0).}

\conference{Joint Proceedings of the ACM IUI Workshops 2023, March 2023, Sydney, Australia}

\title{The Dynamic Creativity of Proto-artifacts in Generative Computational Co-creation}

\author[3]{Juan Salamanca}[%
degree= Ph.D.,
email=jsal@illinois.edu
]
\author[1,2]{Daniel Gómez-Marín}[%
degree= Ph.D.,
email=daniel.gomez@upf.edu
]

\address{Universitat Pompeu Fabra, Music Technology Group.
            Barcelona, Spain}
 \address{Universidad ICESI, Facultad de Ingeniería.
            Cali, Colombia}

\author[1]{Sergi Jordà}[%
degree= Ph.D.,
email=sergi.jorda@upf.edu
]

\address{University of Illinois, School of Art and Design. Urbana-Champaign, United States}

\begin{abstract}
This paper explores the attributes necessary to determine the creative merit of intermediate artifacts produced during a computational co-creative process (CCC) in which a human and an artificial intelligence system collaborate in the generative phase of a creative project. In an active listening experiment, subjects with diverse musical training (N=43) judged unfinished pieces composed by the New Electronic Assistant (NEA). 
The results revealed that a two-attribute definition based on the value and novelty of an artifact (e.g., Corazza’s effectiveness and novelty) suffices to assess unfinished work leading to innovative products, instead of Boden’s classic three-attribute definition of creativity
(value, novelty, and surprise). These findings reduce the creativity metrics needed in CCC processes and simplify the evaluation of the numerous unfinished artifacts generated by computational creative assistants.
\end{abstract}

\begin{keywords}
Computational Co-Creativity \sep creativity assessment \sep dynamic creativity
\end{keywords}

\maketitle


\section{Introduction}
\label{sec:introduction}

\subfile{sections/01_introduction}

\section{Computational Co-Creativity}
\label{sec:Computational_Co_Creativity}
\subfile{sections/02_Comp_Co_creativity}


\subsection{Dynamic Assessment of Computational Co-Creativity}
\label{sec:Dynamic_assessment}
\subfile{sections/04_Dynamic_Assessment}

\section{Computational Co-Creation With The New Electronic Assistant (NEA)}
\label{sec:NEA}
\subfile{sections/05_NEA}

\section{Experiment: Evaluating Creativity In A Computational Co-Creative (CCC) Process}
\label{sec:experiment}
\subfile{sections/12_experiment}

\subsection{Method}
\label{sec:method}
\subfile{sections/13_method}

\subsection{Results}
\label{sec:results}
\subfile{sections/14B_results}

\section{Discussion}
\label{sec:discussion}
\subfile{sections/15_discussion}

\section{Conclusions}
\label{sec:conclusions}
\subfile{sections/16_conclusions}


 \bibliography{cas-refs}

\end{document}

%% file: sections/01_introduction.tex
The increasing learning and predictive capabilities of computational agents consistently find new ways of participating in the arts, design, and humanities, such as in architecture \cite{huang2018architectural}, creative writing \cite{osone2021buncho}, music composition \cite{avdeeff2019artificial}, and video game design \cite{volz2018evolving}, speeding the generation of alternatives, suggesting concepts, generating variants, or automating arrangements. Computational co-creativity (CCC) is the field within the domain of computational creativity (CC) which deals with the \textit{collaborative} process between humans and computational agents aiming at producing creative artifacts. Such collaborative open process in often modeled in terms of sequences of sub-processes, rather than automated generative pipelines  \cite{jordanous2012standardised,colton2011computational}, and many of such models characterize it as iterations over a generative phase and an evaluative phase \cite{lamb2018evaluating}. 

As this type of human-agent collaborative partnership takes hold, the interests of CCC researchers have shifted from studying the creative quality of a system's output (the classic CC approach) to examining how creativity evolves during the ongoing stream of co-creative subprocesses, raising new questions such as: what dimensions should be used to assess a creative process, how should they be interpreted, what information do these dimensions provide about the creative quality of intermediate products, named proto-artifacts. Traces of these dimensions can be found in practitioners' accounts of their experiences with CCC in the arts, spontaneously alluding to concepts traditionally discussed in creativity literature such as surprise: “working with an AI is not dissimilar as working with a human being, given its capacity to surprise you, because that is where the art comes in. That is where the magic comes in, in any kind of performance or working with anybody or anything. [Once surprise comes in] you are able to intervene in what has been generated” \cite{Herndon&dryhurst2021}. 



We argue that the proto-artifacts in human-agent co-creation processes are essential factors of CCC not accounted in traditional CC. Thus, we propose the adoption of Corazza's definition of \textit{dynamic creativity} as it addresses both the process and the potential of proto-artifacts in hybrid collaborative structures. This paper explores the validity of using a two-dimensional definition of creativity (based on originality and effectiveness) \cite{runco2012standard, corazza2016potential}, instead of Boden's three-dimensional one (based on value, novelty and surprise) \cite{boden2010creativity} to assess proto-artifacts generated by computational agents.

The method used is a subject-based empirical evaluation of a musical CCC system that exemplifies the generic CCC process observed in creative disciplines. Subjects with different levels of musical training were integrated into the co-creative workflow of producing an album and appraised the creative quality of intermediate pieces (proto-artifacts). The experimental method takes into account CCC practitioners' reflections on the value of creating with AI systems and relates Boden's \cite{boden2010creativity} and Corazza’s \cite{corazza2016potential} attributes of creativity. The following sections briefly introduce the field of CCC assessment, describe the apparatus used to generate musical pieces, discuss the experiment, and present the results. The paper concludes with reflections about why a two-dimensional assessment of creativity might suffice to characterize CCC processes, as well as its applications and limitations.

%% file: sections/02_Comp_Co_creativity.tex


CCC encompasses creative processes where two or more participants actively collaborate, and at least one of them is a computational agent \cite{jordanous2016four}. While humans adjust the parameters of conceptual spaces, computational agents effectively explore such conceptual spaces reflecting the human ability to select and refine the best ideas \cite{lubart2005can}. As a result, humans and agents make creative, mutually influential contributions to an artefact \cite{davis2013human}. This process coincides with the TOTE behavioral model (test,operate,test,exit) that accounts for how humans execute plans to pursue goals, recursively evaluating the incongruity between the state of machine-generated intermediate artifacts and the intended goal \cite{millerTOTE}. An interesting observation is that a positive affect is often a cue that a person is moving toward the goal and negative affect signals the opposite\cite{CarverScheier1981}. The open ended nature of a CCC process entails that such evaluations are carried out by a human estimator who tries to maximize the chances of achieving something valuable with no certainty of success \cite{corazza2016potential}. In this paper, we are interested in such evaluative dynamics that determine how creative are proto-artifacts in a CCC process.

In traditional CC, generative processes are commonly described as iterative structures 
\cite{wallas1926art,sadler2015wallas,csikszentmihalyi1997flow,simonton2011creativity,ward_smith_finke_1998,amabile2011componential} and can be roughly characterized as two primary sub-processes: producing artifacts and judging them. However, it is not clear how or what to measure in each sub-process. More so if they entail collaborative work in the arts and humanities \cite{yang2020evaluation}. Such indetermination has been a known challenge for CC evaluation frameworks such as FACE \cite{colton2011computational} and SPECS \cite{jordanous2012standardised} because subjects fail to understand the concepts of creativity used in the assessment or because the assessment relies on subjects’ knowledge of the inner workings of the systems being evaluated \cite{lamb2018evaluating}. Nowadays, the question of how to evaluate computational creativity remains open and continues to mature as technical and social applications of systems evolve.


%% file: sections/04_Dynamic_Assessment.tex
A key concern of CCC assessment is the definition of a framework to judge uncompleted work leading to a creative product. Corazza’s definition of \textit {dynamic creativity} is a good starting point: “[c]reativity requires potential originality and effectiveness” \cite [p. 262]{corazza2016potential}. The word ‘potential’ 
conveys the openness of the co-creative process and the latent merit of its intermediate products. A dynamic creative process has a mutable focus, incorporates the intermediate assessment of proto-artifacts, and provides feedback in response to contextual conditions. Corazza suggests that assessing a collaborative creative process implies a dynamic evaluation of an agent’s production of unfinished artifacts by another agent that takes the role of  an estimator. The intermediate outcome of the process is filtered as estimators foresee the consequences of adopting or rejecting such proto-artifacts. Furthermore, Corazza suggests the provocative idea that "[d]iscrepancies between [multiple] estimators’ assessments are a sign of potentially disruptive novelties, generating the necessary energy for transformation of a domain” \cite [p. 265]{corazza2016potential}.



In CC and CCC, estimators (the humans assessing proto-artifacts) are usually required to use two-attribute or three-attribute definitions of creativity as inputs to assess and curate the results of a co-creative process. To mention some of the most prevalent definitions, the \emph{standard definition of creativity} \cite{runco2012standard,stein1953creativity} and the dynamic definition of creativity \cite{corazza2016potential} use originality and effectiveness attributes. Boden defines creativity as the capacity to obtain surprising, valuable, and novel ideas \cite{boden2004creative}. Her definition has been further expanded and formalized mathematically \citep[e.g.][]{WIGGINS2006449}, remaining as the prevalent model in CC literature as it applies to humans or generative devices without preference for either. The following definitions intend to clarify some of these attributes and how they overlap, but are not an exhaustive account of the literature. 

\textbf{Originality}. The originality of an artifact accounts for its authenticity in terms of the independence from precedent realizations, or for the ingenious repositioning of an existing work in an new domain, as does ready-made art. Originality is closely related to novelty because it derives its appraisal from being the first to occupy an unclaimed space in a domain or being the first exemplar of a new domain.

\textbf{Effectiveness}. This term describes the ability to produce a result. It is used in the \emph{standard definition of creativity} as a criterion for eliminating trivial instances that may qualify as original or novel. Some definitions of creativity use usefulness, fit, or appropriateness to convey the same meaning. For Runco and Jaeger \cite{runco2012standard}, it also takes the form of value when creative pieces are appreciated in a market. Corazza uses effectiveness and value interchangeably to convey meaningfulness.

\textbf{Novelty}. The novelty of an artifact can be fully appreciated by the domain-experts as they know a vast portion of the cognitive space of the domain. Therefore, they are responsible for identifying if an artifact extends the domain's boundaries or, even more, transforms the domain itself. For Boden, novelty has two meanings: when something is new to its creator (psychological creativity), and when something comes to life for the first time in human history (historical creativity) \cite{boden2004creative}. Thus, historical creativity corresponds to originality.  

The concept of novelty has been approached scientifically in AI and technology literature, perhaps more than others such as value and surprise \cite{grace2019expectation}. In CC the idea of domain knowledge is implicit in the selection of a training set. That is, novelty is often measured against the training corpus of the AI system, a convenient method for self-assessment of the quality of the generated artifacts. 

\textbf{Surprise}. Assessing surprise during a CCC process gives clues to the level of fulfillment or intensity of the creative process experienced by a subject. 
This attribute has been used to measure creativity in finished artifacts (e.g., \cite{gonzalez2005creativity}), as a synonyms of non-obviousness in patent evaluation \cite{simonton2011creativity}, and as a proxy of the quality of the CCC process from a practitioner’s perspective. Although the subjective nature of surprise could be a shortcoming when judging a finished artifact, it might serve to assess the performance of a computational agent's creativity, especially when the estimator is not a domain expert. Boden specifies three causes of surprise: when unlikely things happen, when unexpected ideas fit know concepts, and when new ideas break the boundaries of established conceptual spaces.  



\textbf{Value}. Creativity researchers generally agree that domain experts classify an artifact as creative when they positively evaluate its value within a field and its subsequent domain. Boden argues that the concept of value, unlike novelty, is elusive \cite{boden2004creative}. The usefulness of value in defining creativity is not straightforward, as social judgments of value change over time, as in the case of artworks that prove to be valuable years after they were first presented, or artists who are considered creative after they have died \cite{Weisberg_2015}. Both value and novelty are subject to the scrutiny of appraisers. They look for virtue in the former dimension, while they look for originality in the latter. The value of a potentially creative artifact cannot be measured until it is deployed and evaluated by estimators. Only domain experts or \emph{gatekeepers} ultimately judge the value of an artifact \cite{csikszentmihalyi1997flow,gluaveanu2015creativity}.

While novelty and surprise are subjects of the cognitive sciences and neuroscience, value is a social construction \cite{heinich2020pragmatic,dewey1939theory}. Corazza claims that "It is, however, evident that novelty and surprise are not disjointed dimensions, because if an item is expected, both surprise and conceptual novelty are denied." \cite[p. 259]{corazza2016potential}. Moreover, recent research claims that novelty and surprise are human reactions independent but close to each other that can be observed and dissociated in electroencephalogram signals. It is suggested that “humans use surprise as a signal to decide when to adapt their behavior, while they use novelty to decide where and what to explore—to eventually develop an improved world-model [...] novelty is more related to memory-recall and surprise is more related to predictions.” \cite [p. 1]{xu2021novelty}. 

In summary, Corazza and Boden propose attribute assessment frameworks with different number of dimensions. Originality and novelty are overlapping concepts specially in relation to Boden's historic creativity. Effectiveness and value refer to the creative purpose from two different view points, for Corazza, effectiveness is defined in terms of practical applications while Boden sees it 
as a social construction. Finally, Corazza acknowledges they are separate concepts but argues that surprise and novelty could be part of a mental process of joint feeling-appreciation.

 None of them can be measured in absolute terms as they are sensitive to how, when, and by who the assessment is made. To the best of our knowledge these frameworks have not been evaluated in practice and this study attempts to shed light on the interplay between them specifically in the assessment of CCC processes. The following sections elaborate and operationalize a generative musical assistant and assess the co-creative process judging the value, novelty and surprise of proto-artifacts in a musical setting. 

%% file: sections/05_NEA.tex
We use the domain of music to exemplify a CCC context in which a computational agent generates real artifacts to be evaluated by human subjects. The complete experiment is presented in Section 4, while this section describes the CCC generative system and its inner workings.

The New Electronic Assistant (NEA) is a music system capable of analyzing a musical style from a symbolic corpus and generating short musical fragments in such style (melody and chord accompaniment). Once a melody and its accompaniment are generated, NEA allows a user to excerpt real-time transformations at four different levels: rhythmic, dynamic, pitch and density\footnote{Basic functionality videos can be found in this link: https://youtube.com/playlist?list=PLD3SOdFCvDNkOBA9Gh3FTAncou-L6xLw1}. 

NEA is designed as a loop generator for music composition and performance. A single NEA instance is suited to complement pre-recorded or real-time performed material or even to conform an ensemble of multiple NEA instances. This latter configuration of instances can be used to achieve rich polyphonic musical arrangements, specially when different NEAs generate complementary melodic styles (e.g., bass lines, main melodies, vocal melodies, etc.) and share information among them (i.e. the chord progression of a generated melody can be transferred to other instances, unifying the whole set of generated melodies). Therefore, provided the performer makes a correct selection of training styles and real-time settings, this parallel propagation of information allows for highly creative mixes of musical material with low effort.

The nature of recent generative interactive systems such as NEA suggests a shift in the traditional workflow of composition and production. Traditionally, a music performance process has required precise embodied coordination in a constant listening/performing cognitive loop, where music is listened to by a performer as she concurrently plays the exact movements on a gesture-to-note instrument contributing to the composition. But interacting with a generative system such as NEA requires a different share of skills. Clicking on a graphical interface of knobs and sliders  replaces the motor skills needed to execute the instrument. The user of NEA seeks to generate a variety of unfinished new melodies in real-time and judge their potential to become something greater until the "right one" is supplied and then transformed. The classic procedure of playing note-by-note in an instrument is replaced by fast critical filtering and real-time parameter tweaking to obtain high-level transformations of intermediate artifacts. There is a shift from motor reaction to fast acoustic discrimination: agile, coordinated embodiment resigns to collected, prospective assessment.

To fulfill the purpose of NEA as a musical generative system, it uses synthesizers and a mixer to convert notes to sound so that the music is perceived. The synthesizers stand out for their ability to model sound in flexible and resourceful ways, especially the sound property known as timbre. Timbre is the character or identity of the source reproducing the notes (i.e., the timbre of a guitar is different than that of a trumpet even though both play the same notes). The mixer, on the other hand, allows control of the intensity of a sound from mute to very loud. At both the synthesizing and mixing stages two complementary systems have been developed allowing for easy prototyping of new material using multiple instances of NEA. Describing these systems is beyond the scope of this paper but in general terms they complete the experience of a NEA's user seeking to automate certain aspects of real-time music creation.


%% file: sections/12_experiment.tex
As presented in the section \ref{sec:Computational_Co_Creativity}, a creative activity can be roughly simplified as a two-stage process: generation of multiple ideas and refinement of the best ones. 
In typical CC experiments, humans or computational agents measure the result of the process, reflecting on how creative the generated output is. Instead, this study is interested in how creative are the proto-artifacts produced during the idea-generation phase.
To that aim, an active listening experiment was designed to examine to what extent subjects apprehend the concepts of novelty, surprise, and value and apply them to assess CCC-generated proto-artifacts. 

Subjects with diverse musical training levels were recruited and contextualized as participants in a musical album production. Their task was to listen to musical stimuli, score them, and decide which ones progress to the next iteration of the creative process. The stimuli were a sample of autonomously generated musical pieces produced by a multi-NEA system. 



%% file: sections/13_method.tex
\textbf{Materials}. Eighteen iterations of the multi-NEA system yielded a constant stream of evolving electronic music pieces scattered throughout the conceptual space of ambient music. One-minute fragments were selected from each of them and used as stimuli. 
In addition, Two randomly selected control stimuli were duplicated to evaluate the consistency of subject’s responses. Specifically, stimuli 1 and 8 are the same, as well as 11 and 19. In total, twenty stimuli were arranged in four different sequences to prevent biases. Each participant listened to one sequence. 

The multi-NEA system used to generate the pieces was trained with classical and pop music melodic styles while timbre and structure were managed by the systems briefly explained in section \ref{sec:NEA}  . 

\textbf{Participants}. The experiment had 43 participants, 37.2\% (16) identified as female, 46.5\% (20) identified as male, and 16.3\% (7) undeclared their gender. Their musical training is homogeneously distributed throughout professional musicians to amateur music producers and performers range. The average musical training is 3.3 (sd = 1.7) on a scale of 1 to 6, with 1 being no-training and 6 being a professional musician. Among participants, 81.4\% (41) have a medium to high knowledge of electronic music, and only 4.6\% (2) reported no knowledge of electronic music.

\textbf{Procedure}. Participants were primed with the following script: For this listening session you are going to play the role of a music producer part of a creative group working on a new album. Your task is to listen to several pieces and assess each so that it continues in the production process or not. The music production process will continue but the essence of the piece will remain close to what you are listening to. Before starting to listen to each of the pieces, the following text was presented: \textit{After listening carefully to this piece please answer: how surprising do you find it? How valuable is it to be published in the album? How novel does it seem to you? Do you have any comments on the piece?} Participants answered the same questions for each of the twenty pieces. For the first three questions, a six-step Likert scale was offered with the following ranges: from  “It is not surprising” to “It is completely surprising”;  from "It is not valuable" to "This piece is very valuable and should be part of the album";  from "It is not a novel piece" to "It is a revolutionary piece". To give a precise sense of the process, subjects were contextualized in an on-going activity. They were unaware the stimuli were made by a machine.

%% file: sections/14B_results.tex
\begin{figure}
    \centering
    \includegraphics[width=0.5\textwidth]{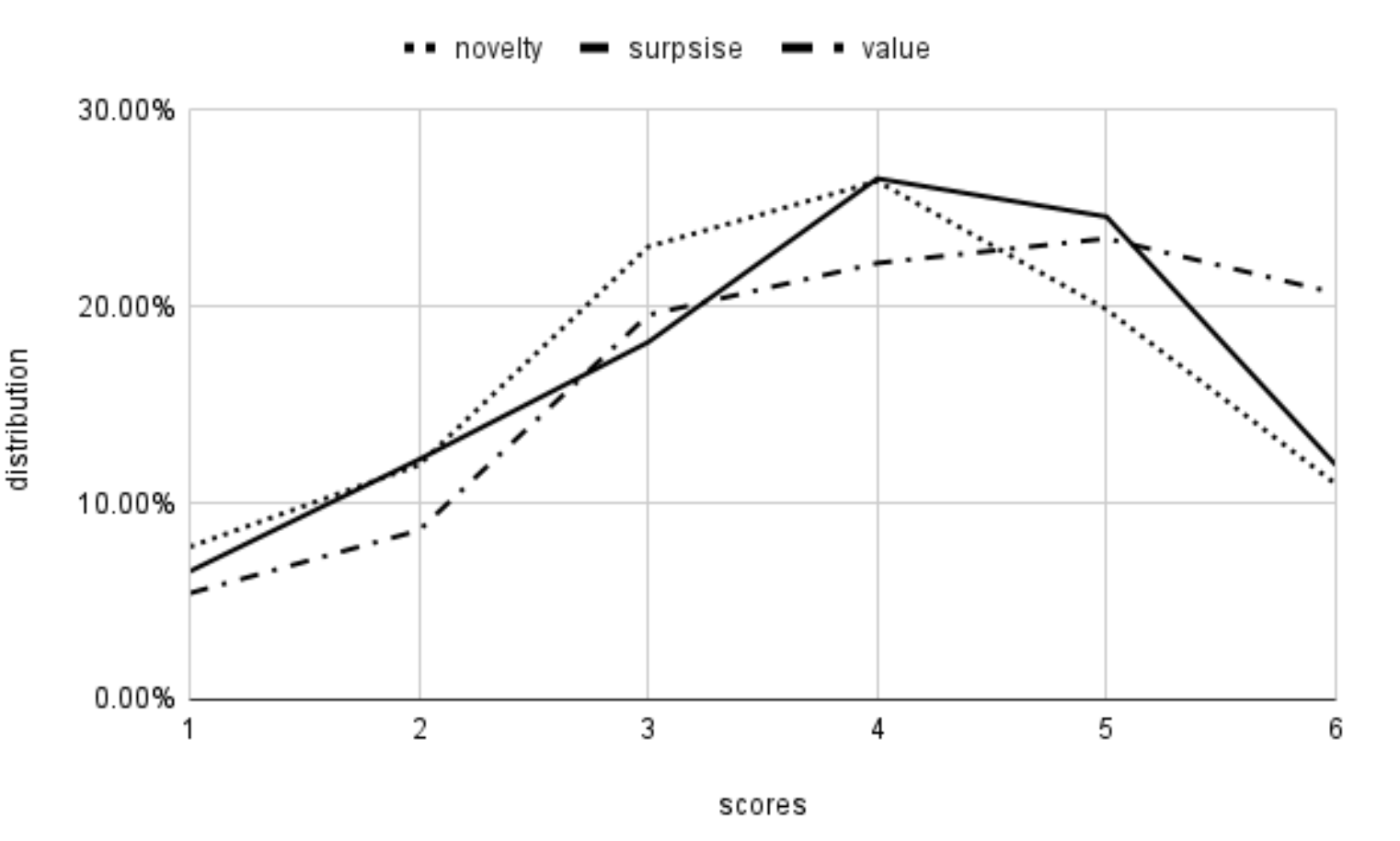}
    \caption{Distribution of novelty, surprise and value scores among all participants and all pieces}
    \label{fig:dist_scores}
\end{figure}

Three subjects with high musical training consistently scored the same control stimuli with a difference of more than 3 points for all three attributes (value, surprise, and novelty). Therefore, all their responses were discarded due to inconsistency.Figure \ref{fig:dist_scores} depicts the distribution of responses. 

\textbf{Discernibility of Boden's three dimensions.} A one-way analysis of variance (ANOVA) carried out to evaluate the difference between the three score sets reveals a statistically significant difference between at least two groups (F(2) = 14.9, p $<$ 0.005). A Tukey’s HSD Test for multiple comparisons found that the mean of value scores (mean = 4.12) was significantly different than the means of novelty and surprise scores (mean = 3.72, p = 0.001, and mean = 3.86 p = 0.0018, respectively). However, there was no statistically significant difference between the means of novelty and surprise scores (p = 0.119).

\textbf{Effect of musical training on creativity scores.} Subjects were segmented in three levels –low, mid, and high– according to their reported musical training to observe if the musical training has a significant effect on creativity scores. For this analysis creativity attributes are considered as treatments and training groups are analyzed as independent categories.

A one-way ANOVA followed by the corresponding Post-Hoc Tukey tests for multiple comparisons revealed that all training levels gave significantly different scores to value and novelty dimensions; only highly trained subjects gave significantly different scores to value and surprise dimensions; and all training levels gave no significant different scores to novelty and surprise dimensions. Moreover, the scores of novelty surprise and value for all pieces by highly trained subjects have higher standard deviations than those of low trained subjects, and those of mid trained subjects (see Table \ref{tab:attribute_pair_training_level}). Echoing \cite{maguire2011making}, results reveal that a highly surprising artifact for an expert might pass as routinary to a novice .

\begin{table*}
\caption{Tukey HSD Post Hoc tests of significance for the difference between attribute ratings for three music training levels (low, mid, and high). Significant values marked with *}
\label{tab:attribute_pair_training_level}

\begin{tabular}{llllllllll}
\hline
                   & \multicolumn{9}{c}{Attribute pair}                                                                            \\ \cline{2-10} 
         & \multicolumn{3}{c}{value-novelty} & \multicolumn{3}{c}{value-surprise} & \multicolumn{3}{c}{novelty-surprise} \\
Training & \multicolumn{1}{c} {mean} & \multicolumn{1}{c} {sd} & \multicolumn{1}{c} {p} & \multicolumn{1}{c} {mean} & \multicolumn{1}{c} {sd} & \multicolumn{1}{c} {p} & \multicolumn{1}{c} {mean} & \multicolumn{1}{c} {sd} & \multicolumn{1}{c} {p} \\  \hline \\
high    & 4.24  & 1.68  & 0.00083*      & 3.61  & 1.63  & 0.01663*      & 3.76  & 1.62  & 0.64244   \\
mid     & 4.34  & 1.23  & 0.00243*      & 3.99  & 1.07  & 0.08534       & 4.11  & 1.07  & 0.43779   \\
low     & 3.87  & 1.44  & 0.0229137*    & 3.56  & 1.44  & 0.4016408     & 3.73  & 1.46  & 0.3700743                           
\end{tabular}%
\end{table*}

\begin{figure*}[ht]
\centering
\includegraphics[width=1\textwidth]{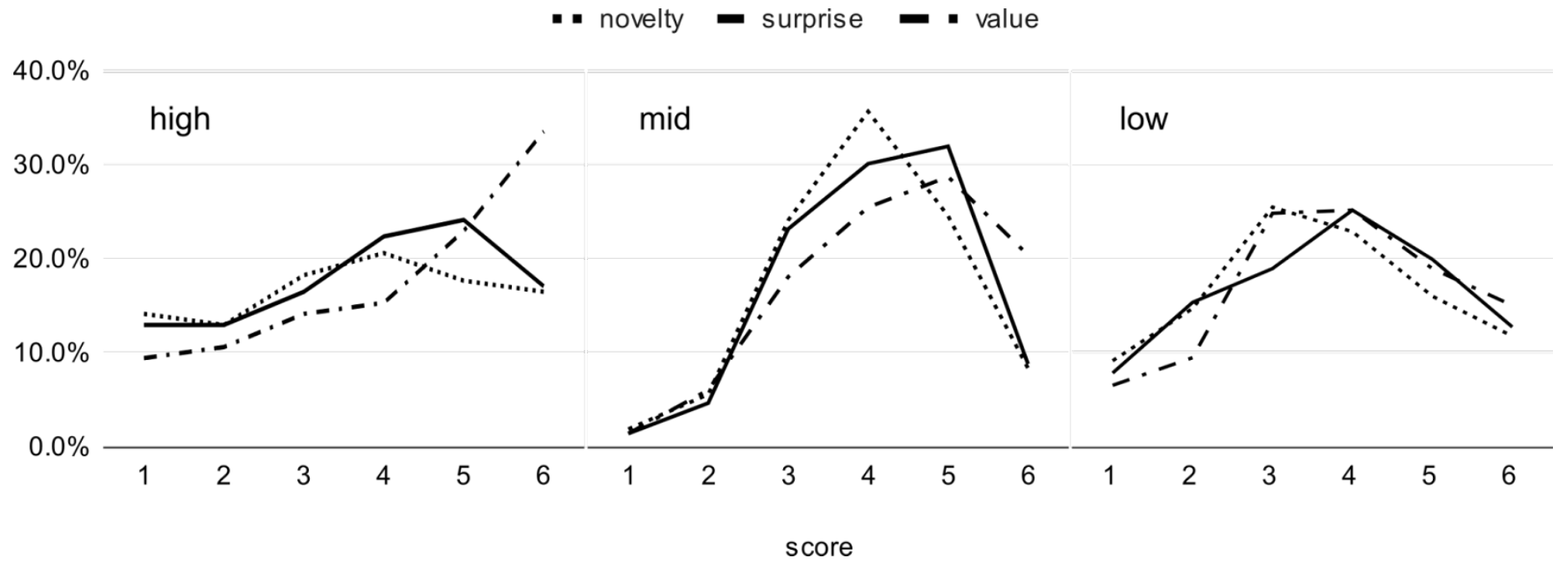}
\caption{Novelty, surprise and value at three different musical training levels: high, mid and low (left, bottom, right respectively)}
\label{fig:training_effect}
\end{figure*}

\textbf{Similarity of creativity scores across training groups.} A complementary analysis was carried out having musical training as treatments and creativity attributes as categories (see Figure \ref{fig:training_effect} and Table \ref{tab:training_level_attribute_score}). This serves to discern to what extent the training level refines estimator's creativity appraisal.

A series of one-way ANOVAs, one for each creativity attribute, followed by corresponding Tukey’s HSD Test for multiple comparisons showed that the scores of mid and low trained subjects are significantly different for all three attributes. High and low trained subjects have significantly different value scores, while high and mid trained groups have significantly different novelty scores (see \ref{tab:training_level_attribute_score}). The dispersion of novelty and surprise scores are very similar for all subject segments, but value scores are more dispersed than those of novelty and surprise in high and mid trained subjects (sd= 1.679 vs 1.608, 1.607 and sd= 1.242 vs 1.092, 1.081 respectively). In the case of low-trained subjects the opposite effects is observed: value scores are less disperse than those of novelty and surprise (sd= 1.406 vs 1.449, 1.460). 

The effect of training in value scores has proven to be statistically significant between mid and low training. The analysis reveals that the standard deviation of scores of highly trained subjects is significantly greater than the ones of the rest of the subjects.

\begin{table*}[]
\caption{Tukey HSD post hoc tests of significance for the difference between high, mid, and low levels of musical training.  Significant differences marked with *}
\label{tab:training_level_attribute_score}
\begin{tabular}{lccc}
\hline
          & \multicolumn{3}{c}{Training level} \\ \cline{2-4} 
Attribute & High-Mid   & High-Low   & Mid-Low  \\ \hline
Novelty   & 0.04*      & 0.89       & 0.004*   \\
Surprise  & 0.101      & 0.722      & 0.005*   \\
Value     & 0.789      & 0.019*     & 0.001*  
\end{tabular}%
\end{table*}

%% file: sections/15_discussion.tex
The results from the statistical tests elucidate whether a three-attribute definition of creativity (value, surprise, and novelty) accounts for proto-artifacts creativity in the generative phase of co-creative processes and how to interpret such metrics. While value scores are significantly different from surprise and novelty scores, surprise and novelty ones are close to each other. Further comparative analysis between pairs of scores reveals three clear insights: value stands out as a different concept from novelty, novelty and surprise appear as non-discernible concepts \footnote{This is compliant with \cite{xu2021novelty} as they suggest surprise and novelty are cognitive processes that operate closely.}, and value and surprise appear discernible to highly trained subjects but mid and low trained subjects have similar mental constructs for value and surprise. Consequently, one could cautiously argue that two-attribute models of creativity could suffice expert estimators to assess unfinished artifacts during a co-creative process. 

\textbf{Is potential creativity a two or three-attribute space?} The empirical results obtained show that novelty and surprise responses are not statistically distinguishable, suggesting that these attributes, although different in meaning, have a joint appraisal in experimental conditions. This resonates with the two-dimensional co-creativity assessment models proposed by the standard definition of creativity, and Kantosalo et al. (value and novelty, plus the quality of user interaction) \cite{kantosalo2020modalities}. Consequently, in evaluating  proto-artifacts during a co-creative process a two-attribute model of value and originality could account for Boden's three-attribute model.

Reducing the dimension of the attribute space while preserving the assessment quality simplifies human or agent estimator’s tasks. Such reduction has practical implications in multiple real-life scenarios that require filtering large sets of artifacts created during CCC processes. Indeed, as CCC processes permeate human creative activities, the number and quality of potentially creative artifacts that need assessment will most likely grow exponentially, demanding effective and adequate metrics to carry out estimation tasks. To assess the creative potential of creating with such computational assistants, one would need to measure the creative quality of proto-artifacts produced during the idea generation phase.

However, one could argue that the observed proximity between the novelty and surprise concepts can result from the experimental conditions. On the relation between experiencing surprise and rating novelty, Xu et al.  explain how “humans use surprise as a signal to decide when to adapt their behavior, while they use novelty to decide where and what to explore—to eventually develop an improved world-model.” \cite[p.1]{xu2021novelty} This idea suggests that both attributes are used in conjunction to adjust expectations dynamically. They operate independently yet contribute to broader cognitive processing. It is necessary to investigate whether the closeness of these concepts stems from the unfinished nature of stimuli that confounds their subjective assessment or from the training level of estimators participating in the study.

\textbf{The effect of domain training in assessing proto-artifacts.} There is a plausible effect of domain knowledge in scores of the three creativity attributes. 
The higher the training the greater the significance of the differences between value and surprise, and value and novelty (see Table \ref{tab:attribute_pair_training_level} columns 4 and 7). But the inverse effect is observed between value and surprise: the higher the training the lower the significance between novelty and surprise (see Table \ref{tab:attribute_pair_training_level} column 10). This evidence shows that as training becomes more specialized, subjects are more confident gauging value, yet they learn that not every valuable artifact is surprising.
In particular, highly trained subjects encounter more pieces with extreme value scores than mid or low trained subjects (see Figure \ref{fig:training_effect}). That is, experts used the whole semantic range of the evaluation scale, while non-experts concentrate their scores around the second third. A triangulation of Tukey Post Hoc test of effects for experts reinforces the claim that novelty and surprise are not discernible, while value is the only attribute with statistically significant difference between high and low trained subjects. This suggests that training has a higher positive effect on the ability to appreciate value than novelty or to experience surprise. In other words, domain expertise is especially expressed when assessing value and not so much when assessing novelty or surprise. A potential explanation is that training builds a more nuanced domain-specific cognition and reinforces the estimator’s capacity to determine the value of unfinished artifacts in terms of the foreseen potential to evolve into more refined pieces or branch out novel variations worth exploring.  



%% file: sections/16_conclusions.tex
This paper argues for the adoption of a dynamic framework to judge uncompleted work (deemed proto-artifacts) leading to a creative product in the context of computational co-creation (CCC) processes. Such approach derived from Corazza's dynamic definition of creativity, recognizes that artists engaged in computational co-creation not only estimate the creative merit of their work once the 
piece is finished, but assess the creative potential of intermediate proto-artifacts at each iteration of the generative process. Intermediate assessments depict how a CCC process may go about and put forward the potential anticipation of creative outcomes from the early stages. Hence, a suitable computational assistant should maximize the creative potential of the process, either by enhancing the human's generative capacity or by facilitating recurrent proto-artifacts assessments.

The findings of an active listening experiment conducted to determine the creative quality of unfinished musical pieces generated by NEA (New Electronic Assistant) suggest that in an experimental setting subjects' appraisal of novelty and surprise is not discernible. Thus, a two-attributes definition of creativity could account for Boden's three-attributes definition. Even though novelty and surprise represent different creative attributes, originality could account for both of them because novelty and surprise tend to blend in subjective assessments of creativity, while value is certainly differentiable, especially for domain experts. 

For the time being, a two dimensional creativity assessment of proto-artifacts is not invalidated, and may simplify assessment procedures with subjects.  We suggest using the dimensions of value and originality (rather than Corazzas' effectiveness and originality). Value is preferred to effectiveness because it conveys meaningfulness in a variety of fields, including the arts, better than the functional notion of effectiveness. On the other hand, the responses of subjects with three levels of expertise in the domain studied showed that novelty and surprise are two different but coupled mental operations. The former is related to memory and the ability to forget and the latter is related to the stability of short-term predictions. This suggests that the assessment of one could be a proxy for the other. For practical research purposes, it makes more sense to use fewer dimensions to conduct large-scale experiments, especially with lay subjects for whom these concepts generally remain fuzzy.



Finally, as AI permeates human creative activities of all sorts the generation of proto-creative material flourishes. That is, an unavoidable bi-product of assisted creativity is the proliferation of unfinished artifacts that must be assessed not only by humans but also by AI agents. Such increase in potentially creative outcomes calls out for the implementation of assertive assessment methods. The results presented here might prove useful to define further methodologies for effective human and agent-based assessment of creative artifacts in CCC scenarios.